\documentclass[12pt]{article}
\usepackage{amsfonts, amsmath, amssymb, amsthm, graphicx, latexsym, xy, verbatim, bm, hyperref}

\usepackage{setspace}

\newcommand{\m}{\mu}
\newcommand{\cc}{\Lambda}

\newcommand{\bnabla}{\bar{\nabla}}

\newcommand{\be}{\begin{equation}}
\newcommand{\ee}{\end{equation}}

\title{Kundt spacetimes of massive gravity in three dimensions}

\author{Mohamed Chakhad\\
{\small \textit{Theory Group, Department of Physics,
University of Texas at Austin,
Austin, TX 78712, USA}}}

\begin{document}

\begin{flushright}
UTTG-07-07\\
July 2009
\end{flushright}
\vskip 0.2in
\begin{center}
{\LARGE Kundt spacetimes of massive gravity in three dimensions} \vskip 0.2in
{\large Mohamed Chakhad}\\
\vskip 0.1in
\textit{Theory Group, Department of Physics,\\
University of Texas at Austin,\\
Austin, TX 78712, USA\\}
\vskip 0.1in
{\small email: chakhad@physics.utexas.edu}\\
\end{center}

\vskip 0.3in

\begin{abstract}
We study Kundt solutions of topologically massive gravity (TMG) and the new theory of massive gravity (NMG), proposed recently in \cite{Bergshoeff:2009hq}. For topologically massive gravity, only the CSI Kundt solutions (i.e., solutions with constant scalar polynomial curvature invariants) were found very recently in \cite{Chow}. We find non-CSI explicit solutions of TMG, when $\cc=-\m^2$, and these are the first such solutions. For the new theory of massive gravity, after reducing the field equations to a manageable system of differential equations, the CSI solutions are discussed in detail, with a focus on a subfamily whose solutions are particulary easy to describe. A number of properties of Kundt solutions of TMG and NMG, such as an identification of solutions which lie at the intersection of the full non-linear and linearized theories, are also derived.
\end{abstract}

\section{Introduction}

In the search for consistent quantum theories of gravity, a great amount of attention has recently been devoted to gravity in three dimensions, spurred in part by the papers of Witten \cite{Witten:2007kt}, of Li, Song, and Strominger \cite{Li:2008a}, and of Bergshoeff, Hohm, and Townsend \cite{Bergshoeff:2009hq}.
While Witten's paper is an attempt at an exact solution of Einstein gravity on AdS$_3$, the last two papers consider theories of massive gravity. In \cite{Li:2008a}, topologically massive gravity (TMG) \cite{DJT} with Brown--Henneaux boundary conditions \cite{Brown:1986nw}, at a special value of the couplings, was argued to be stable and classically chiral, and was dubbed chiral gravity. More recently, a new theory of massive gravity in three dimensions (NMG) has been proposed in \cite{Bergshoeff:2009hq}, and since its inception, a number of papers have appeared which address the same considerations for NMG as those that have appeared before for TMG.

In recent developments in TMG, the debate around ``logarithmic'' solutions of TMG \cite{log solutions}, which appear at the chiral point, has culminated in the discovery of new consistent asymptotic boundary conditions \cite{Henneaux:2009}, which are less restrictive than the old Brown--Henneaux boundary conditions, and thus log gravity  \cite{Maloney:2009} was born. On another front, away from the chiral point, solutions of TMG other than AdS have been proposed as stable vacua of the theory \cite{Anninos:2008fx}. Despite the surge of interest in three-dimensional theories of massive gravity, there still remains a lot to be learned about the classical space of solutions of both theories of massive gravity. In light of the preceding remarks and for its own interest, a scan of the space of classical solutions across all values of the parameters appearing in the equations of motion of massive gravity in three dimensions and their classification is worthwhile. This paper is a contribution to this search effort.

In three spacetime dimensions the Riemann curvature tensor is determined algebraically by the Ricci tensor, making all vacuum solutions of Einstein's equation locally equivalent, and implying the absence of local propagating degrees of freedom. This is not the case for topologically massive gravity and the new theory of massive gravity, which admit solutions with non-constant curvature. In fact, we will find explicit solutions of TMG with non-constant scalar polynomial curvature invariants.

The nonlinearity of the equations of motion makes the task of finding their solutions difficult. Thus one resorts to simplifying assumptions, by restricting one's attention to a specific class of spacetimes. The classes can be defined either in terms of an \emph{ad hoc} ansatz, or more satisfactorily via (for example) admitted symmetries or the existence of special, not necessarily Killing, vector fields. Kundt's class of spacetimes \cite{Kundt:1961} belongs to the latter category, since it is defined by the existence of a geodesic null vector with vanishing optical scalars. The Kundt class serves as an umbrella under which to unify many of the solutions of TMG and NMG which have appeared in the literature. It is also host to explicit solutions which were not found before.

In this paper, we study the local problem of which Kundt metrics are solutions of the field equations of massive gravity in three dimensions. This problem, in the context of topologically massive gravity, was studied very recently in \cite{Chow}, where only the solutions with constant scalar polynomial curvature invariants were found explicitly. After collecting some preliminaries in Section 1, we study Kundt solutions of TMG in Section 2. We first provide an adapted review of the results of \cite{Chow}, which serves as a framework in which to discuss the new solutions of subsection 2.2 and facilitates a comparison with the results for the new theory of massive gravity. In subsection 2.3, after making a choice of background metrics, we address the question of which Kundt metrics are solutions to both the full non-linear and linearized equations of motion of TMG. Section 3 is devoted to a study of Kundt solutions of NMG. After simplifying the field equations, we briefly discuss the special case $\lambda=m^2$ in subsection 3.1. We then discuss in detail the CSI solutions in subsection 3.2, with emphasis on an easily describable subfamily. In subsection 3.3, with an appropriate choice of parameters, we answer a question about which Kundt metrics are exact solutions to both TMG and NMG. Finally, in subsection 3.4, we identify solutions of the full non-linear equation of motion of NMG which are also solutions of the linearized theory, with the same choice of background metrics as for TMG.

\section{Preliminaries}

\subsection{Topologically massive gravity}

The action for topologically massive gravity \cite{DJT} is given by
$$I_{\rm TMG}= \frac{1}{16\pi G}\Big(I_{\rm EH}+\frac{1}{\m}I_{\rm CS}\Big),$$
where the Einstein--Hilbert action, $I_{\rm EH}$, and the gravitational Chern--Simons action, $I_{\rm CS}$, are
\begin{eqnarray*}
I_{\rm EH}&=&\int d^3x \sqrt{-g}\big(R-2\cc\big),\\
I_{\rm CS}&=&\frac{1}{2}\int
d^3x\sqrt{-g}\epsilon^{\lambda\mu\nu}\Gamma^\rho_{\lambda\sigma}
\Big(\partial_\mu\Gamma^\sigma_{\rho\nu}+\frac{2}{3}\Gamma^\sigma_{\mu\tau}\Gamma^\tau_{\nu\rho}\Big).
\end{eqnarray*}
The mass parameter $\m$ is a non-zero real number, $g$ denotes the determinant of the metric, and $\epsilon^{\lambda\mu\nu}=\varepsilon^{\lambda\mu\nu}/\sqrt{-g}$, with $\varepsilon^{012}=1$.

Variation of the TMG action with respect to the metric yields the equation of motion
\be
\label{TMG}
R_{\alpha\beta}-\frac{1}{2}Rg_{\alpha\beta}+\cc g_{\alpha\beta}+\frac{1}{\m}C_{\alpha\beta}=0,
\ee
where $C_{\alpha\beta}$ is the Cotton tensor, given by
$$C_{\alpha\beta}=\epsilon_\alpha{}^{\mu\nu}\nabla_\mu
\Big(R_{\nu\beta}-\frac{1}{4}R g_{\nu\beta}\Big).$$

In three dimensions, conformal flatness is characterized by the vanishing of the Cotton tensor\footnote{For manifolds of dimension less than four, the Weyl tensor vanishes identically.}. Thus, there are two broad classes of solutions to the field equation of TMG (\ref{TMG}): conformally flat metrics---which are necessarily solutions to the Einstein equation---and metrics with non-zero Cotton tensor. The Cotton tensor is (i) traceless, (ii)  covariantly conserved, and (iii) symmetric (due to the the Bianchi identity). Furthermore, under a parity transformation, the Cotton tensor changes sign. This parity breaking entails that a solution of (\ref{TMG}) consists of a metric and an orientation of the manifold. However, note that given a solution of (\ref{TMG}) for $\m=\m_0$, the same metric with a flip of the orientation of the manifold solves (\ref{TMG}) for $\m=-\m_0$. Finally, by virtue of the tracelessness of the Cotton tensor, a solution of (\ref{TMG}) is necessarily of constant scalar curvature: $R=6\cc$.

\subsection{The new theory of massive gravity}

The action of the new theory of massive gravity \cite{Bergshoeff:2009hq} is given by
$$I_{\rm NMG}=\frac{1}{16\pi G}\int d^3x\sqrt{-g}\Big(R-2\lambda -\frac{1}{m^2}K\Big),$$
where the scalar $K$ is a higher derivative term:
$$K=R^{\mu\nu}R_{\mu\nu}-\frac{3}{8}R^2.$$
The constant $\lambda$ is not necessarily equal to the cosmological constant $\cc$, and $m$ is a non-zero mass parameter, which (to be inclusive) is allowed to be either real or pure imaginary, so that $m^2$ is a non-zero real number of either sign.
The equation of motion is
\be
\label{NMG}
R_{\alpha\beta}-\frac{1}{2}Rg_{\alpha\beta}+\lambda g_{\alpha\beta}-\frac{1}{2m^2}K_{\alpha\beta}=0,
\ee
where the tensor $K_{\alpha\beta}$, given by
\begin{eqnarray}
\nonumber K_{\alpha\beta}&=&2\nabla^2R_{\alpha\beta}-\frac{1}{2}\Big(\nabla_\alpha\nabla_\beta R+g_{\alpha\beta}\nabla^2R\Big)-8R^\mu_\alpha R_{\mu\beta}\\
\nonumber&&+\frac{9}{2}RR_{\alpha\beta}+\Big(3R^{\mu\nu}R_{\mu\nu}-\frac{13}{8}R^2\Big)g_{\alpha\beta},
\end{eqnarray}
is covariantly constant and satisfies the important property: $g^{\alpha\beta}K_{\alpha\beta}=K$. The trace of the equation of motion (\ref{NMG}) gives $K+m^2(R-6\lambda)=0$, so that, for a solution of (\ref{NMG}), $K=0$ if and only if $R=6\lambda$. For an Einstein metric, with $R_{\alpha\beta}=2\cc g_{\alpha\beta}$, to be a solution of (\ref{NMG}), it is necessary and sufficient that
$$\lambda=\cc-\frac{\cc^2}{4m^2}.$$
This relation implies that equation (\ref{NMG}) admits an Einstein solution if and only if $\lambda/m^2\leq1$. For positive $m^2$, anti-de~Sitter spacetime is a solution of (\ref{NMG}) if and only if $\lambda/m^2<0$; Minkowski spacetime is a solution if and only if $\lambda=0$; but de-Sitter spacetime is a solution for any $\lambda/m^2\leq1$. For negative $m^2$, the same statements hold as for positive $m^2$, with AdS and dS interchanged.

\subsection{Linearized equations of motion}

To obtain the equation of motion for the linearized excitations around a constant-curvature background metric $\bar{g}_{\alpha\beta}$, with Ricci tensor $\bar{R}_{\alpha\beta}=2\cc\bar{g}_{\alpha\beta}$, the full metric is written as
$$g_{\alpha\beta}=\bar{g}_{\alpha\beta}+\epsilon h_{\alpha\beta},$$
and the different terms of the non-linear equation of motion are expanded up to first order in the parameter $\epsilon$. The coefficients of $\epsilon$ in the expansion of the Ricci tensor, scalar curvature, Cotton tensor, and $K_{\alpha\beta}$ are respectively given by \cite{Deser:2003,Olmez:2005,Liu:2009}
\begin{eqnarray*}
R^{(1)}_{\alpha\beta}&=&\frac{1}{2}\big(-\bnabla^2h_{\alpha\beta}-\bnabla_\alpha\bnabla_\beta h+2\bnabla^\mu\bnabla_{(\alpha}h_{\beta)\mu}\big),\\
R^{(1)}&=&-\bnabla^2h+\bnabla^\mu\bnabla^\nu h_{\mu\nu}-2\cc h,\\
C^{(1)}_{\alpha\beta}&=&\bar{\varepsilon}_\alpha{}^{\mu\nu}
\bnabla_\mu\Big(R^{(1)}_{\nu\beta}-\frac{1}{4}\bar{g}_{\nu\beta}R^{(1)}-2\cc h_{\nu\beta}\Big),\\
K^{(1)}_{\alpha\beta}&=&-\frac{1}{2}\bnabla^2R^{(1)}\bar{g}_{\alpha\beta}-\frac{1}{2}\bnabla_\alpha\bnabla_\beta R^{(1)}+2\bnabla^2R^{(1)}_{\alpha\beta}-4\cc\bnabla^2h_{\alpha\beta}\\
&&-5\cc R^{(1)}_{\alpha\beta}+\frac{3}{2}\cc R^{(1)}\bar{g}_{\alpha\beta}+\frac{19}{2}\cc^2h_{\alpha\beta},
\end{eqnarray*}
where the indices are raised and lowered using the background metric, the barred quantities are constructed from the background metric, and $h=\bar{g}^{\alpha\beta}h_{\alpha\beta}$.

The linearized equation of motion for TMG (\ref{TMG}) is
$$R^{(1)}_{\alpha\beta}-\frac{1}{2}R^{(1)}\bar{g}_{\alpha\beta}-2\cc h_{\alpha\beta}+\frac{1}{\m}C^{(1)}_{\alpha\beta}=0.$$
Taking the trace of this equation with respect to the background metric, we find that $h_{\alpha\beta}$ is a solution only if
$$R^{(1)}=0.$$
Thus, for a fluctuation $h_{\alpha\beta}$ which satisfies this necessary condition, the equation of motion becomes
\be \label{LTMG}
\Big(R^{(1)}_{\alpha\beta}-2\cc h_{\alpha\beta}\Big)+\frac{1}{\m}\bar{\varepsilon}_\alpha{}^{\mu\nu}
\bnabla_\mu\Big(R^{(1)}_{\nu\beta}-2\cc h_{\nu\beta}\Big)=0.
\ee

Similarly, the linearized equation of motion for NMG (\ref{NMG}) is
\be \label{LNMG}
R^{(1)}_{\alpha\beta}-\frac{1}{2}R^{(1)}\bar{g}_{\alpha\beta}+(\lambda-3\cc) h_{\alpha\beta}-\frac{1}{2m^2}K^{(1)}_{\alpha\beta}=0.
\ee
Taking the trace of this equation, and using $\lambda=\cc -\cc^2/(4m^2)$, we obtain
$$\left(1-\frac{\cc}{2m^2}\right)R^{(1)}=0,$$
which is equivalent to $R^{(1)}=0$ if $\cc\neq 2m^2$ (or $\lambda\neq m^2$). Thus, for $\cc\neq 2m^2$, the linearized equation of motion for NMG becomes
$$\left(1+\frac{5\cc}{2m^2}\right)\big[R^{(1)}_{\alpha\beta}-2\cc h_{\alpha\beta}\big]-\frac{1}{m^2}\bnabla^2\big[R^{(1)}_{\alpha\beta}-2\cc h_{\alpha\beta}\big]=0.$$

\subsection{Segre types of the Ricci and Cotton tensors}

In order to place our solutions of massive gravity in the broader context of three-dimensional spacetimes and (in doing so) provide some of their properties, we shall give a brief review of the algebraic classification of symmetric tensors in Lorentzian manifolds. Let $(M,g_{\alpha\beta})$ be a three-dimesional Lorentzian manifold, let $p\in M$, and let $S_{\alpha\beta}$ be a symmetric tensor at $p$. The eigenvalue problem
\be
\nonumber S^\alpha_\beta v^\beta=\lambda v^\alpha
\ee
(where $S^\alpha_\beta=g^{\alpha\sigma}S_{\sigma\beta}$,  $\lambda$ is a complex number, and $v^\alpha$ is a tangent vector at $p$) gives rise to a classification of $S_{\alpha\beta}$ in terms of the Jordan canonical form of $S^\alpha_\beta$ and a coarser classification by Segre type \cite{Hall:1987bz,Santos,Hall:1999}. The classification via Segre type reveals only the dimension of the Jordan blocks and the degeneracy of the eigenvalues, where equal eigenvalues are enclosed in parentheses. For example, $S_{\alpha\beta}$ is of Segre type \{(21)\} if: the Jordan canonical form of $S^\alpha_\beta$ contains two blocks of dimension 2 and 1, and the two eigenvalues are equal. In the case of non-real eigenvalues, the Segre type is denoted \{1$z\bar{z}$\}. All Segre types (i.e., types \{111\}, \{1$z\bar{z}$\}, \{21\}, \{3\}, and their degeneracies) are possible, but the fact that $S_{\alpha\beta}$ is symmetric places restrictions on the possible eigenvectors of $S^\alpha_\beta$. For instance, (i) a timelike eigenvector is possible only for type \{111\} and its degeneracies, in which case it is unique. By convention, a comma is used to indicate the presence of a timelike eigenvector, and most importantly, to distinguish between the two types \{(1,1)1\} and \{1,(11)\}. (ii) The types \{21\}, \{(21)\}, and \{3\} admit exactly one null eigenvector, and this eigenvector corresponds to the Jordan block of dimension 2 in the first two types. (iii) The type \{(1,1)1\} admits exactly two null eigenvectors, and it is the only type with this property. (iv) All types admit a spacelike eigenvector, except for types \{(1,1)1\} and \{3\}. Finally, (v) the types \{1,11\}, \{1,(11)\}, and \{1$z\bar{z}$\} do not admit null eigenvectors.

The above algebraic classification holds for any symmetric tensor of either type (0,2) or (2,0). In particular, we are interested in the Segre types of the tensors appearing in the equations of motion of massive gravity, with focus on the Segre types of the Ricci and Cotton tensors. As is evident from the vacuum equation of motion of topologically massive gravity (\ref{TMG}), the Ricci and Cotton tensor for a solution of (\ref{TMG}) have the same Segre type and the same eigenvectors, at any given point of the spacetime. The tracelessness of the Cotton tensor implies that the eigenvalues of the Ricci tensor, for solutions of (\ref{TMG}), are constrained to have their sum equal to $6\cc$. Moreover, a solution of (\ref{TMG}) with Ricci tensor of type \{(111)\} in a neighborhood is necessarily conformally flat in that neighborhood. As for the vacuum solutions of the new theory of massive gravity, a glance at equation (\ref{NMG}) shows that the tensor $K_{\alpha\beta}$ and the Ricci tensor have the same Segre type and same eigenvectors.

\subsection{Kundt spacetimes}

In four dimensions, important transverse properties of a null geodesic vector field  can be expressed in terms of three scalars, called the optical scalars: the expansion (or divergence), the shear, and the twist. A physical interpretation of these scalars in terms of the shadow cast on a screen by an opaque object, with both object and screen orthogonal to the null geodesic vector, was given by Sachs in \cite{Sachs} .

Spacetimes which admit a null vector field with vanishing optical scalars were first considered as a class, in the context of General Relativity, by Wolfgang Kundt \cite{Kundt:1961}. This important class of spacetimes, dubbed Kundt's class, is the subject of chapter 31 of \cite{Stephani:2003}. Being defined in invariant terms, the Kundt class lends itself to a straightforward generalization to any number of dimensions. Thus, a Lorentzian manifold, of dimension $D\geq3$, is called a \emph{Kundt spacetime} if it admits a geodesic null vector with zero twist, zero shear, and zero divergence. In $D$ dimensions, the optical scalars of a null geodesic vector $n^\alpha$ are given by \cite{Chow}:
$$\theta=\frac{1}{D-2}\nabla_\alpha n^\alpha,\;\;\;\;\sigma^2=\nabla^\alpha n^\beta\nabla_{(\alpha}n_{\beta)}-\frac{1}{D-2}\big(\nabla_\alpha n^\alpha\big)^2,\;\;\;\;\omega^2=\nabla^\alpha n^\beta\nabla_{[\alpha}n_{\beta]},$$
where $\theta$, $\sigma^2$, and $\omega^2$ correspond to the expansion, shear, and twist, respectively.
In three dimensions, every null vector field is automatically twist-free and shear-free, essentially because Sachs' ``screen'' is one-dimensional, so three-dimensional Kundt spacetimes are those spacetimes which admit a divergence-free null geodesic vector field. As argued (for example) in \cite{CSI}, a D-dimensional Kundt spacetime admits a local adapted coordinate system, $(u,v,x^i)$, in which the metric has the following form
$$ ds^2=2du(Hdu+dv+W_idx^i)+\tilde{g}_{ij}(u,x^k)dx^idx^j,$$
where $H=H(u,v,x^k)$, $W_i=W_i(u,v,x^k)$, and $i,j,k=1,\ldots,D-2$. It is easy to check that the vector $n^\alpha\partial_\alpha=\partial_v$ is null, geodesic (and affinely parameterized), and divergence-free.

In three dimensions, let $x=x^1$, $W=W_1$, and $\tilde{g}=\tilde{g}_{11}$. Then, using the coordinate transformation $x\rightarrow\int \sqrt{\tilde{g}(u,x)}dx$, we can set $\tilde{g}(u,x)=1$, so that
\be \label{Kundt}
ds^2=2du(Hdu+dv+Wdx)+dx^2.
\ee
From this line element, we see that while the coordinate $x$ is spacelike and the coordinate $v$ is null, the causal character of the coordinate $u$ depends on the function $H$. As a side remark, the family of metrics given by (\ref{Kundt}) is closed under the action of the Kerr--Schild transformation $g_{\alpha\beta}\rightarrow g_{\alpha\beta}+\phi n_\alpha n_\beta$, where $\phi$ is a function of $u$, $v$, and $x$. Indeed, this Kerr--Schild transformation generates the entire family of metrics (\ref{Kundt}), starting from the subfamily of metrics with vanishing $H$.

It is worth mentioning that the Kundt class is host to a large diversity of spacetimes. For instance, the Kundt metric $ds^2=2du(u xdu+dv+v^2dx)+dx^2$ admits no local symmetries. However, a number of properties of the spacetimes given by (\ref{Kundt}) follow at once by an examination of the nowhere-vanishing vector fields in the (special) direction of $n^\alpha\partial_\alpha=\partial_v$, i.e., the vector fields $f\partial_v$, where $f$ is a nowhere-vanishing function of $u$, $v$, and $x$. (i) $f\partial_v$ is recurrent\footnote{A vector field $v^\alpha$ is said to be \emph{recurrent} (or parallel) if $\nabla_\alpha v_\beta=R_\alpha v_\beta$, for some vector field $R^\alpha$, called the \emph{recurrence vector} (\cite{Stephani:2003}, page 69).} if and only if $\partial_v W=0$; the recurrence vector of $f \partial_v$ is in the direction of $\partial_v$ if and only if $\partial_v f=\partial_x f=0$. (ii) $f \partial_v$ is covariantly constant for some $f$ if and only if $\partial_v W=0$ and $\partial_v H$ depends only on $u$, in which case $f \partial_v$ is covariantly constant if and only if $f=\exp(-\int \partial_v Hdu)$. (iii) $f \partial_v$ is a Killing vector for some $f$ if and only if $\partial_v^2 W=\partial_v^2 H=0$ and $\partial_u\partial_v W=\partial_x\partial_v H$; $f \partial_v$ is a conformal Killing vector if and only if it is a Killing vector. (iv) $R_{\alpha\beta}n^\alpha n^\beta=0$ everywhere. (v) $C_{\alpha\beta}n^\alpha n^\beta=0$ if and only if $\partial_v^3 W=0$. (vi) $\partial_v$ is an eigendirection of the Ricci tensor if and only if $\partial_v^2 W=0$. (vii) $\partial_v$ is an eigendirection of the Cotton tensor if and only if $\partial_v^3 W=\partial_v^3 H=0$.

A spacetime is said to be \emph{CSI} \cite{CSI} if all scalar polynomial curvature invariants are constant. (The term ``polynomial'' is crucial in this definition, since there are CSI spacetimes which admit other types of non-vanishing local scalar invariants \cite{Page}.) For the Kundt metric (\ref{Kundt}) to be CSI, it is necessary and sufficient \cite{CSI3D} that there exist constants $c_1$, $c_2$, and $c_3$ such that
\be
\label{CSIcondition1}
W=W_0+vW_1\;\;\;\; {\rm and}\;\;\;\; H=H_0+vH_1+\frac{1}{8}v^2(W_1^2+4c_1),
\ee
where $W_0$, $W_1$, $H_0$, and $H_1$ are functions of $u$ and $x$, and $W_1$ satisfies the equations
\be
\label{CSIcondition2}
\partial_x W_1-\frac{1}{2}W_1^2=2c_2\;\;\;\; {\rm and}\;\;\;\; (c_1-c_2)W_1=c_3.
\ee
Furthermore, the CSI Kundt metrics with $c_1=c_2$ have Ricci and Cotton tensors of Segre type \{3\}, \{(21)\}, or \{(111)\}; we call these collectively \emph{type A}. For $c_1\neq c_2$ and $c_2\neq 0$, the Ricci and Cotton tensors have Segre type \{21\} or \{(1,1)1\}; we call these CSI Kundt metrics \emph{type B}. Finally, for $c_2=0$ and $c_1\neq0$, the Ricci tensor has Segre type \{21\} or \{(1,1)1\}, but the Cotton tensor has type \{3\}, \{(21)\}, or \{(111)\}; we call these CSI Kundt metrics \emph{type C}. Note that, at a given point, the Ricci and Cotton tensors for CSI Kundt metrics of types A and B do not necessarily have the same Segre type.

The definitions of the last paragraph lead immediately to the following consequences. CSI Kundt metrics of type B cannot be conformally flat. A CSI Kundt metrics of type C cannot be a  solution of Einstein's equation. CSI Kundt solutions of the vacuum field equation of TMG, equation (\ref{TMG}), cannot be of type C. However, as we will see, type C solutions of NMG do exist, but only for $\lambda=m^2$.

\section{Solutions of topologically massive gravity}

In this section, after reducing the field equation of topologically massive gravity (\ref{TMG}) for the Kundt family of metrics (\ref{Kundt}), we will first give an adapted review of the CSI solutions, which were found very recently in \cite{Chow}. The emphasis will be on a subfamily of CSI solutions, whose description is particularly easy. Then, new explicit Kundt solutions for $\cc=-\m^2$ will be given. These are the first explicit non-CSI solutions of equation (\ref{TMG}). Finally, with a choice of suitable backgrounds, we will identify Kundt metrics which are solutions of both the full and linearized equations of motion of TMG.

Let $E_{\alpha\beta}$ denote the TMG tensor, i.e.,
$$E_{\alpha\beta}=R_{\alpha\beta}-\frac{1}{2}Rg_{\alpha\beta}+\cc g_{\alpha\beta}+\frac{1}{\m}C_{\alpha\beta}.$$
By direct calculation in the adapted coordinate system $(u,v,x)$, where the Kundt metric takes the form (\ref{Kundt}), we find that the equations $E_{vv}=0$, $E_{vx}=0$, and $\partial_v^2R=0$ (where $R$ is the scalar curvature) are satisfied if and only if
$$W=W_0+vW_1\;\;\;\; {\rm and}\;\;\;\; H=H_0+vH_1+v^2H_2,$$
where $W_0$, $W_1$, $H_0$, $H_1$, and $H_2$ are all functions of $u$ and $x$. Since $\partial_v^2W$ is identically zero, the null vector $\partial_v$ is everywhere an eigendirection of the Ricci tensor, and consequently, the Ricci tensor cannot be anywhere of Segre type \{1,11\}, \{1,(11)\}, or \{1$z\bar{z}$\}.

The function $W_0$ can be eliminated by the coordinate transformation $v\rightarrow v+f(u,x)$, where the function $f$ satisfies the equation $\partial_xf+W_1f=W_0$. Incidentally, $W_1$ and $H_2$ are not changed by this transformation.

Now, the scalar curvature is
$$R=4H_2+2\partial_xW_1-\frac{3}{2}W_1^2.$$
Since all vacuum solutions of topologically massive gravity have $R=6\cc$, the function $H_2$ is then given by
\be
\label{H2TMG}
H_2=-\frac{1}{2}\partial_xW_1+\frac{3}{8}W_1^2+\frac{3}{2}\cc.
\ee

Using the necessary conditions found thus far, and calculating the components $E_{uv}$ and $E_{xx}$, one obtains $E_{uv}=-2E_{xx}$, and the vanishing of these components gives the nonlinear second-order differential equation
\be
\label{Exx}
\frac{2}{3}\partial_x\big[\partial_x W_1-\frac{1}{2}W_1^2-2\cc\big]-\Big(W_1-\frac{2}{3}\m\Big)\big[\partial_x W_1-\frac{1}{2}W_1^2-2\cc\big]=0.
\ee
Since this equation is independent of the variable $u$ and is translationally symmetric with respect to the variable $x$, the solution takes the form $W_1=W_1(x+f(u))$. Now, a coordinate transformation of the form $x\rightarrow x+f(u)$ allows us to eliminate the dependence of $W_1$ on $u$. Thus, equation (\ref{Exx}) is an ODE in $W_1$.  Note that the equation is written in this form to make manifest the CSI solutions, which will be discussed shortly.

The vanishing of the remaining components of the TMG tensor provides two more differential equations, restricting the functions $H_1$ and $H_0$. More specifically, the vanishing of $E_{ux}$ leads to the differential equation
\be
\label{Eux}
2\partial_x^2H_1-(W_1-2\m)\partial_x H_1=0.
\ee
Given a solution of equation (\ref{Exx}), equation (\ref{Eux}) is then (for fixed $u$) a second-order homogeneous linear ODE in $H_1$. The vanishing of $E_{uu}$ yields the differential equation
\begin{eqnarray}
\label {Euu}
\nonumber 0&=&4\partial_x^3 H_0+2\big[3W_1+2\m\big]\partial_x^2 H_0+\big[14\partial_x W_1-W_1^2+4\m W_1-12\cc\big]\partial_x H_0\\
&& +\,2\big[4\partial_x^2 W_1-(W_1-2\m)\partial_x W_1\big]H_0+4\big[\partial_u\partial_x H_1+H_1\partial_x H_1\big].
\end{eqnarray}
Given a solution $W_1$ of equation (\ref{Exx}) and a solution $H_1$ of equation (\ref{Eux}), equation (\ref{Euu}) is then essentially a third-order non-homogeneous linear ODE, with coefficients depending on $W_1$ and the non-homogeneous term depending on $H_1$.

In summary, a Kundt solution of (\ref{TMG}) necessarily admits a coordinate system, $(u,v,x)$, in which the metric takes the form
\be
\label{Kundt2}
ds^2=2du\big[(H_0+vH_1+v^2H_2)du+dv+vW_1dx\big]+dx^2,
\ee
where $W_1$ and $H_2$ are functions of $x$, and $H_1$ and $H_0$ are functions of $u$ and $x$. This family of metrics will be the setting for many of our later considerations. For a Kundt metric of the form (\ref{Kundt2}) to be a vacuum solution of topologically massive gravity, the functions $W_1$, $H_2$, $H_1$, and $H_0$ must satisfy equations (\ref{H2TMG})--(\ref{Euu}). The conformally flat solutions of TMG satisfy the following equations
$$\partial_xW_1-\frac{1}{2}W_1^2=2\cc,\;\;\;\;H_2=\frac{1}{8}\big(W_1^2+4\cc\big),$$
$$\partial_xH_1=0,\;\;\;\;2\partial_x^2H_0+2W_1\partial_xH_0+\big(W_1^2+4\cc\big)H_0=0.$$

The square of the Ricci tensor of a vacuum solution of TMG of the form (\ref{Kundt2}) is given by
\be
\label{sqRicci}
R_{\alpha\beta}R^{\alpha\beta}=18\cc^2+\frac{3}{2}\Big(\partial_xW_1
-\frac{1}{2}W_1^2-4\cc\Big)\Big(\partial_xW_1-\frac{1}{2}W_1^2\Big),
\ee
which is clearly constant if and only if $\partial_xW_1-W_1^2/2$ is constant. In what follows, we first discuss the CSI solutions, which were given in \cite{Chow}, after which we will present new solutions with non-constant $R_{\alpha\beta}R^{\alpha\beta}$.

\subsection{CSI Kundt solutions of TMG}

\begin{figure}[thp]
\centering
\includegraphics[width=\textwidth]{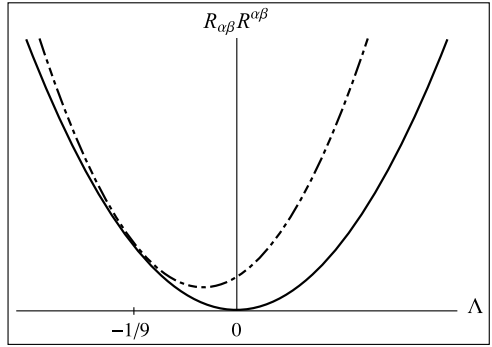}
\caption{The square of the Ricci tensor (in units of $\m^4$) of the CSI Kundt solutions of TMG versus $\cc$ (in units of $\m^2$). The solid curves correspond to metrics of type A and the dashed curves to metrics of type B.}\label{TMGfig1}
\end{figure}

Recall that the conditions for a Kundt metric to be CSI are given by (\ref{CSIcondition1}) and (\ref{CSIcondition2}). Combining these requirements with equations (\ref{H2TMG}) and (\ref{Exx}), it is easy to see that the CSI Kundt solutions of TMG must satisfy
$$\partial_x W_1-\frac{1}{2}W_1^2=2\cc\;\;\;\; {\rm or}\;\;\;\; W_1=\frac{2}{3}\m.$$
While the first condition leads to CSI Kundt solutions of type A (which are defined in the discussion following equation (\ref{CSIcondition2})), the second condition leads to CSI Kundt solutions of type B, except when $\cc=-\m^2/9$, in which case there are only type A solutions. The CSI Kundt solutions of equation (\ref{TMG}) are given in explicit form in \cite{Chow}, so we will content ourselves here with plotting the square of the Ricci tensor versus the cosmological constant for these solutions, since this plot illustrates the special nature of the point $\cc=-\m^2/9$, where the two types of solutions merge. For type A solutions, the square of the Ricci tensor is
$$R_{\alpha\beta}R^{\alpha\beta}=12\cc^2,$$
while for type B, it is
$$R_{\alpha\beta}R^{\alpha\beta}=\frac{2}{27}\big(243\cc^2+18\m^2\cc+\m^4\big).$$
The square of the Ricci tensor (in units of $\m^4$) is plotted against $\cc$ (in units of $\m^2$) in figure \ref{TMGfig1}. The two curves for type A and type B meet tangentially at  $\cc=-\m^2/9$, and the type B solutions disappear at this point. While the square of the Ricci tensor of the type A solutions admits $0$ as a global minimum at $\cc=0$, the square of the Ricci tensor for type B solutions admits a minimum at $\cc=-\m^2/27$, with value $R_{\alpha\beta}R^{\alpha\beta}=4\m^4/81$.

In the remainder of this subsection, we shall examine the solutions of TMG with constant $W_1$. These solutions are necessarily CSI, and $W_1$ is given by either $W_1^2=-4\cc$ (solutions of type A) or $W_1=2\m/3$ (solutions of type B). The solutions with $W_1^2=-4\cc$ are obviously possible only for $\cc\leq 0$. Part of the reason for examining these solutions is that the differential equations in $H_1$ and $H_0$ (equations (\ref{Eux}) and (\ref{Euu}), respectively) are (for fixed $u$) linear ODEs with constant coefficients. Hence their solutions are completely determined by the roots of their respective characteristic polynomials and the non-homogeneous term in the case of equation (\ref{Euu}). The exact form of these solutions is, again, given in \cite{Chow}, so we will focus here on the multiplicity of the roots of the characteristic polynomials and whether or not they are real. While non-real roots yield solutions which are oscillatory in the $x$ direction, the interest in roots with multiplicity greater than one stems from the fact that their appearance at the chiral point $\cc=-\m^2$ gives rise to interesting solutions of TMG satisfying the boundary conditions of log gravity \cite{Henneaux:2009}.

The roots of the characteristic polynomial of equation (\ref{Eux}) are $0$ and $W_1/2-\m$. The two roots are real, and they become a double root if $W_1=2\m$, which happens if and only if $W_1^2=-4\cc$ and $\cc=-\m^2$. This means that the roots of the characteristic polynomial of equation (\ref{Eux}) are distinct for all solutions of type B.

Now, the characteristic equation of the linear ODE in $H_0$, equation (\ref{Euu}), is
\be
\label{cEuu}
r\big[4r^2+2(3W_1+2\m)r-W_1^2+4\m W_1-12\cc\big]=0.
\ee
For solutions with $W_1^2=-4\cc$ (which are solutions of type A), the roots are $0$, $-W_1$, and $-W_1/2-\m$. The discriminant of equation (\ref{cEuu}), in this case, is (up to a positive factor) equal to $-\cc(\cc+\m^2)^2$. Since the discriminant is non-negative, all roots are real. Moreover, the characteristic polynomial admits a multiple root if $\cc=0$ or $\cc=-\m^2$. At $\cc=0$, $0$ is a double root. For $\cc=-\m^2$, $0$ is also a double root when $W_1=-2\m$, but $-2\m$ is the double root if $W_1=2\m$.

\begin{figure}[htp]
\centering
\includegraphics[width=\textwidth]{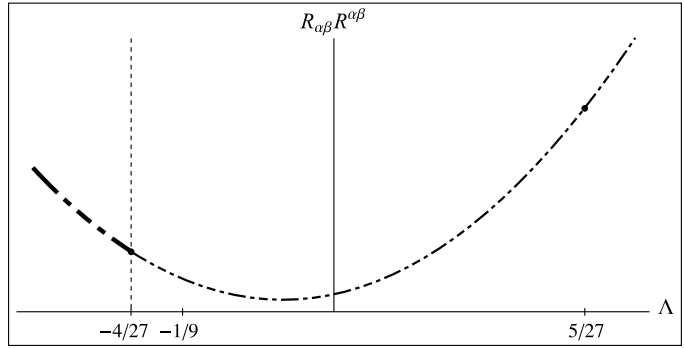}
\caption{Nature of the roots of equation (\ref{cEuu}) for the CSI Kundt solutions of TMG of type B. The presence of non-real roots is indicated by the thick dashed curve. Multiple roots occur at $\cc=-4\m^2/27$ and $\cc=5\m^2/27$. $R^{\alpha\beta}R_{\alpha\beta}$ and $\cc$ are expressed in units of $\m^4$ and $\m^2$ respectively.}\label{TMGfig2}
\end{figure}

Finally, for solutions with $W_1=2\m/3$ (which are the solutions of type B), the roots of equation (\ref{cEuu}) are $0$ and $-\m\pm\sqrt{27\cc+4\m^2}/3$. The discriminant of equation (\ref{cEuu}) is (up to a positive factor) equal to $(27\cc+4\m^2)(27\cc-5\m^2)^2$. Since this discriminant is negative for $\cc<-4\m^2/27$, equation (\ref{cEuu}) admits two non-real (complex conjugate) roots in this interval (see figure \ref{TMGfig2}). The roots are all real for $\cc\geq\-4\m^2/27$. There are multiple roots at $\cc=-4\m^2/27$ and $\cc=5\m^2/27$. At $\cc=-4\m^2/27$, $-\m$ is a double root, and at $\cc=5\m^2/27$, $0$ is a double root.

\subsection{Non-CSI solutions of TMG}

In our search for non-CSI solutions of topologically massive gravity, we must clearly look for solutions of equation (\ref{Exx}) such that $\partial_xW_1-W_1^2/2$ is not constant. Equation (\ref{Exx}) is an autonomous second-order ordinary differential equation. Interestingly, the equation admits a one-parameter group of Lie point symmetries (generated by the obvious translational symmetry with respect to the independent variable), unless $\cc=-\m^2$, in which case the symmetry algebra is two-dimensional. This means that for $\cc=-\m^2$, by Lie's theory of differential equations, the ODE reduces to two quadratures.

\begin{figure}[thp]
\centering
\includegraphics[width=\textwidth]{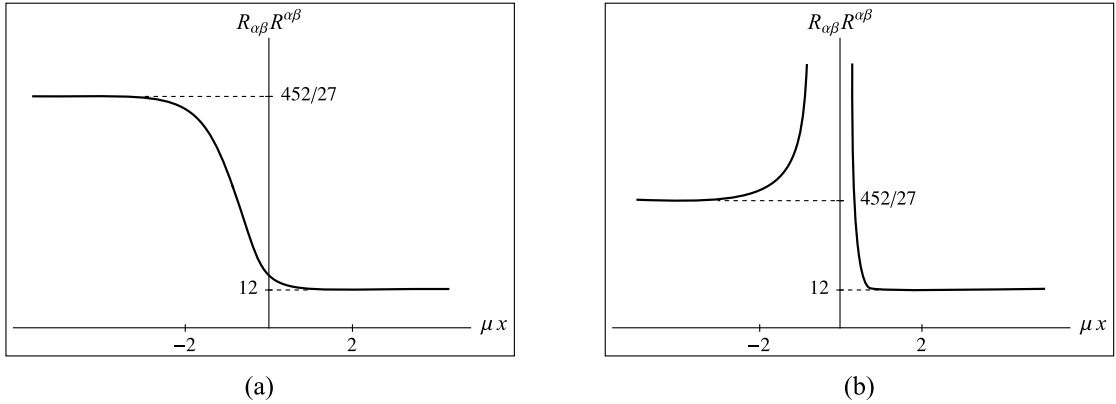}
\caption{The square of the Ricci tensor (in units of $\m^4$) of (a) the new Kundt solutions of TMG satisfying (\ref{new1}) and (b) the new Kundt solutions satisfying (\ref{new2}).}\label{new}
\end{figure}

The infinitesimal generators of the Lie point symmetries for equation (\ref{Exx}), in the special case when $\cc=-\m^2$, are $\partial_x$ and $\exp(2\m x)[\partial_x-2\m (W_1+2\m)\partial_{W_1}]$. We first transform equation (\ref{Exx}) using the change of independent and dependent variables $\m z=\exp(-2\m x)$, $w=\exp(2\m x)(W_1+2\m)$. The resulting equation is simpler in form than equation (\ref{Exx}):
\be
\label{chiral}
\frac{d^2w}{dz^2}+\frac{5}{4} w\frac{dw}{dz}+\frac{3}{16}w^3=0.
\ee
This equation belongs to the family of equations $y''+ayy'+by^3$, which was studied recently in \cite{Chandra}. It is also worth noting that the general solution to this equation was given, long ago, in Kamke's book \cite{Kamke}, page 551. While the general solution can be written in implicit form, explicit solutions of equation (\ref{chiral}) can be found, as revealed from the first integral
$$a\left(\frac{dw}{dz}+\frac{1}{4}w^2\right)^2
+b\left(\frac{dw}{dz}+\frac{3}{8}w^2\right)^3=0,$$
where $a$ and $b$ are two constants. While the choice $b=0$ (and $a\neq0$) leads to CSI solutions, the choice $a=0$ (and $b\neq0$) leads to the first-order differential equation
$$\partial_xW_1-\frac{3}{4}W_1^2-\m W_1+\m^2=0,$$
with solutions
\be
\label{new1}
W_1=-\frac{2\m}{3}\big[2\tanh(\m x)+1\big],
\ee
and
\be
\label{new2}
W_1=-\frac{2\m}{3}\big[2\coth(\m x)+1\big].
\ee
Note that the constant of integration is eliminated by a coordinate transformation of the form $x\rightarrow x+c$. The square of the Ricci tensor for these new families of Kundt solutions of TMG is
$$R_{\alpha\beta}R^{\alpha\beta}=12\cc^2+\frac{128\cc^2}{27}\big(e^{2\m x}\pm 1\big)^{-4},$$
where the upper sign corresponds to the solutions satisfying (\ref{new1}), and the lower sign corresponds to the solutions satisfying (\ref{new2}). A plot of the square of the Ricci tensor as a function of $x$ is given in figure \ref{new}. While the coordinate $x$ spans the whole real line for the solutions satisfying (\ref{new1}), it is restricted to either positive or negative values for the solutions satisfying (\ref{new2}).

Now, since equations (\ref{Eux}) and (\ref{Euu}) are satisfied if $H_1=0$ and $H_0=0$, then the metrics
$$ds^2=-\frac{8\m^2v^2}{3\big(e^{2\m x}\pm 1\big)}du^2+2dudv-\frac{4\m v\big(3e^{2\m x}\mp 1\big)}{3\big(e^{2\m x}\pm 1\big)}dudx+dx^2$$
are non-CSI solutions of topologically massive gravity (\ref{TMG}) for $\cc=-\m^2$. We content ourselves here with providing a short list of their properties. The Segre type of the Ricci tensor of these metrics is generically \{21\} and degenerates to \{(1,1)1\} on the Killing horizon $v=0$. The corresponding eigenvalues are $2\cc+(8\cc/9)[\exp(2\m x)\pm 1]^{-2}$  and $2\cc-(16\cc/9)[\exp(2\m x)\pm 1]^{-2}$, respectively. The algebra of infinitesimal symmetries is the two-dimensional non-abelian Lie algebra generated by the Killing vectors $\partial_u$ and $u\partial_u-v\partial_v$. Finally, note that the solution with the upper sign interpolates between the AdS metric (in the limit $x\rightarrow +\infty$), which is a CSI metric of type A, and a CSI metric of type B (in the limit $x\rightarrow -\infty$). As for the solution with the lower sign, the square of the Ricci tensor blows up in the limit $x\rightarrow 0$, tends to that of AdS spacetime in the limit $\m x\rightarrow +\infty$, and tends to that of a CSI Kundt solution of type B in the limit $\m x\rightarrow -\infty$.

\subsection{Kundt solutions of TMG which are also solutions of linearized TMG}

We shall now characterize, among the family of Kundt metrics (\ref{Kundt2}), those metrics which are simultaneously solutions of the full non-linear vacuum equation of motion of topologically massive gravity (\ref{TMG}) and solutions of the linearized equation of motion (\ref{LTMG}), with background metric
\be
\label{background}
\bar{g}_{\alpha\beta}dx^\alpha dx^\beta=2\left[\frac{1}{8}v^2\big(\bar{W}_1^2+4\cc\big)du^2+dudv
+v\bar{W}_1dudx\right]+dx^2,
\ee
where the function $\bar{W}_1(x)$ is a solution of the equation $$\partial_x\bar{W}_1-\frac{1}{2}\bar{W}_1^2=2\cc.$$
The background metric (\ref{background}) is locally AdS, Minkowski, or dS, depending on whether $\cc$ is negative, zero, or positive, respectively. In the Minkowski case, the coordinate system $(u,v,x)$ is a global coordinate system. The chosen backgrounds (\ref{background}) are characterized, among the family of metrics given by (\ref{Kundt}), as those metrics which have constant curvature and admit the non-abelian two-dimensional algebra generated by the vector fields $\partial_u$ and $u\partial_u-v\partial_v$ as algebra of infinitesimal symmetries.

Let's assume the Kundt metric given by (\ref{Kundt2}) is a solution of the full equation of motion (\ref{TMG}). This means that the functions $H_2$, $W_1$, $H_1$, and $H_0$ satisfy equations (\ref{H2TMG})--(\ref{Euu}). In what follows we will find necessary and sufficient conditions for such a metric to be a solution of the linearized equation of motion as well.

Computing $R^{(1)}$ yields
$$R^{(1)}=\frac{3}{2}\big(W_1-\bar{W}_1\big)^2.$$
The necessary vanishing of $R^{(1)}$ then implies that we must have $W_1=\bar{W}_1$. Consequently, $H_2$ takes the simple form
$$H_2=\frac{1}{8}\big(\bar{W}_1^2+4\cc\big).$$

Now let $L_{\alpha\beta}$ denote the left-hand side of the linearized equation of motion (\ref{LTMG}):
$$L_{\alpha\beta}=\big(R^{(1)}_{\alpha\beta}-2\cc h_{\alpha\beta}\big)+\frac{1}{\m}\bar{\varepsilon}_\alpha{}^{\mu\nu}
\bnabla_\mu\big(R^{(1)}_{\nu\beta}-2\cc h_{\nu\beta}\big).$$
A direct computation of $L_{\alpha\beta}$ shows that all components are identically zero (modulo the full nonlinear field equations), except for $L_{uu}$, which gives the additional equation
$$H_1\partial_xH_1=0.$$
This equation and equation (\ref{Eux}) are simultaneously satisfied  if and only if $\partial_xH_1=0$.

In summary, the Kundt metric (\ref{Kundt2}) is a solution of both the full and the linearized equation of motion of topologically massive gravity, with the background metric (\ref{background}), if and only if the functions $W_1$, $H_2$, $H_1$, and $H_0$ satisfy the following equations
$$W_1=\bar{W}_1,\;\;\;\; H_2=\frac{1}{8}\big(\bar{W}_1^2+4\cc\big),\;\;\;\;\partial_xH_1=0,$$
and equation (\ref{Euu}). Note that these solutions include non-conformally flat metrics.

\section{Solutions of the new massive gravity}

In this section, we study solutions of the vacuum equation of motion of the new theory of massive gravity (\ref{NMG}) among the Kundt family of metrics given by (\ref{Kundt}). Let $\tilde{E}_{\alpha\beta}$ denote the NMG tensor, i.e., $$\tilde{E}_{\alpha\beta}=R_{\alpha\beta}-\frac{1}{2}Rg_{\alpha\beta}
+\lambda g_{\alpha\beta}
-\frac{1}{2m^2}K_{\alpha\beta}.$$
By direct calculation, the components of the NMG tensor with the simplest expressions are $\tilde{E}_{vv}$, $\tilde{E}_{vx}$, $\tilde{E}_{uv}$, and $\tilde{E}_{xx}$. However, as can be checked, there is a relation between these four components and the trace of the NMG tensor:
$$2\tilde{E}_{uv}=g^{\alpha\beta}\tilde{E}_{\alpha\beta}
-\tilde{E}_{xx}+\big(2H-W^2\big)\tilde{E}_{vv}+2W\tilde{E}_{vx}.$$ This identity allows us to exchange the equation $\tilde{E}_{uv}=0$ for the simpler equation $g^{\alpha\beta}\tilde{E}_{\alpha\beta}=0$. For ease of notation, we denote the trace of the NMG tensor by $\tilde{T}$, i.e.,
$$\tilde{T}=g^{\alpha\beta}\tilde{E}_{\alpha\beta}.$$

The two components $\tilde{E}_{vv}$ and $\tilde{E}_{vx}$ are given by
\begin{eqnarray}
\nonumber 4m^2\tilde{E}_{vv}&=&2\partial_v^4H-W\partial_v^4W-
3(\partial_vW)\partial_v^3W+(\partial_v^2W)^2+2\partial_x\partial_v^3W,\\
\nonumber 16m^2\tilde{E}_{vx}&=&16m^2W\tilde{E}_{vv}-8W\partial_v^4H
-20(\partial_vW)\partial_v^3H-4(\partial_v^2W)\partial_v^2H\\
\nonumber && +\,32(\partial_v^3W)\partial_vH+16(\partial_v^4W)H+(\partial_vW)^2\partial_v^2W\\
\nonumber &&+\,8\partial_x\partial_v^3H-16\partial_u\partial_v^3W+8m^2\partial_v^2W.
\end{eqnarray}
Using the two equations $\tilde{E}_{vv}=0$ and $\tilde{E}_{vx}=0$, and \emph{assuming} $W$ is a polynomial function in $v$, we show that $W$ is necessarily linear in $v$. This assumption and the ensuing result will then lead to a manageable system of equations. Let $W$ be a polynomial function in $v$ of degree $n\geq 2$, i.e., $W=W_0+\cdots+v^nW_n$, where the coefficients $W_i$ ($i=0,\ldots,n$) are functions of $u$ and $x$. Using this ansatz for $W$ and computing the $(2n-4)$th derivative of $\tilde{E}_{vv}$ with respect to $v$, we find that Kundt solutions of NMG must satisfy the following simple identity
$$\partial_v^{2n}H=a_nW_n^2,$$
where $a_n=(2n-4)!n(n-1)(4n^2-15n+12)/2$. Since $W_n$ does not depend on $v$, the last identity implies that all derivatives of $H$ with respect to $v$ higher than $2n$ must be identically zero. Now, if we calculate the $(3n-4)$th derivative of $\tilde{E}_{vx}$ with respect to $v$, we obtain a second identity for the $(2n)$th derivative of $H$ with respect to $v$:
$$b_nW_n\partial_v^{2n}H+c_nW_n^3=0,$$
where $b_n=(3n-4)!n(n-1)(4n^2+5n-4)/(2n)!$ and $c_n=(3n-4)!n(n-1)(5n^2-20n+16)/8$. Combining the two identities, we obtain the necessary condition
$$(a_nb_n+c_n)W_n^3=0.$$
For $n\geq 3$, $a_n$, $b_n$, and $c_n$ are all positive, and for $n=2$, we have $a_2b_2+c_2=-28/3$. This means that the factor $(a_nb_n+c_n)$ is not zero for any $n\geq 2$, and consequently $W_n$ must be identically zero. Thus the function $W$ is necessarily linear in $v$, i.e.,
$$W=W_0+vW_1,$$
where $W_0$ and $W_1$ are functions of $u$ and $x$.

This result simplifies the expressions for $\tilde{E}_{vv}$, $\tilde{E}_{vx}$, and $\partial_v^2\tilde{T}$ drastically:
\begin{eqnarray*}
2m^2\tilde{E}_{vv}&=&\partial_v^4H,\\
4m^2\tilde{E}_{vx}&=&2\partial_x\partial_v^3H
-5W_1\partial_v^3H,\\
-8m^2\partial_v^2\tilde{T}&=&
4(\partial_v^3H)^2+\big(4\partial_v^2H-4\partial_xW_1+W_1^2+8m^2\big)
\partial_v^4H.
\end{eqnarray*}
It is easy to see that the three equations $\tilde{E}_{vv}=0$, $\tilde{E}_{vx}=0$, and $\partial_v^2\tilde{T}=0$ are simultaneously satisfied if and only if the function $H$ is quadratic in $v$, i.e.,
$$H=H_0+vH_1+v^2H_2,$$
where $H_0$, $H_1$, and $H_2$ are functions of $u$ and $x$. Thus, a Kundt metric (\ref{Kundt}), with $W$ polynomial in $v$, is a solution of the vacuum equation of motion of the new theory of massive gravity (\ref{NMG}) only if $W$ is linear in $v$ and $H$ is quadratic in $v$. This immediately implies that the null vector $\partial_v$ is everywhere an eigendirection of both the Ricci and Cotton tensors. Also, as before, the function $W_0$ is eliminated by a coordinate transformation of the form $v\rightarrow v+f(u,x)$.

There are four remaining equations that the Kundt metric (\ref{Kundt}) must satisfy in order to be a vacuum solution of the new theory of massive gravity.  The equations $\tilde{T}=0$ and $\tilde{E}_{xx}=0$ are coupled differential equations of the functions $H_2$ and $W_1$, containing derivatives with respect to $x$ only. The vanishing of $\tilde{T}$ gives the equation
\be
\label{T}
0=\big(8H_2-W_1^2-8m^2\big)\big(8H_2-8\partial_xW_1
+3W_1^2+24m^2\big)-192m^2\big(\lambda-m^2\big),
\ee
whereas the vanishing of $\tilde{E}_{xx}$ yields the equation
\begin{eqnarray}
\label{tExx}
\nonumber 0&=&128\partial_x^2H_2-320W_1\partial_xH_2+64H_2^2
-16\big(8\partial_xW_1-11W_1^2+8m^2\big)H_2\\
&&+\,W_1^4+16m^2W_1^2+64\lambda m^2.
\end{eqnarray}
Since these equations are independent of the variable $u$ and are translationally symmetric with respect to the variable $x$, the solutions take the form $W_1=W_1(x+f(u))$ and $H_2=H_2(x+f(u))$. Thus, a coordinate transformation of the form $x\rightarrow x+f(u)$, allows us to to eliminate the $u$ dependence of $W_1$ and $H_2$.

Now, assuming all the necessary conditions derived so far are satisfied, equation $\tilde{E}_{ux}=0$ is (for constant $u$) a third-order homogeneous linear ODE in $H_1$, where the coefficients depend on $H_2$ and $W_1$:
\be
\label{tEux}
0=8\partial_x^3H_1-8W_1\partial_x^2H_1+\big(8H_2 -4\partial_xW_1+W_1^2-8m^2\big)\partial_xH_1,
\ee

Upon solving equation (\ref{tEux}) for $H_1$, the equation $\tilde{E}_{uu}=0$ is (for constant $u$) a fourth-order non-homogeneous linear ODE in $H_0$, where the coefficients of the homogeneous part depend on $H_2$ and $W_1$, and the non-homogeneous term depends on $H_1$:
\begin{eqnarray}
\label{tEuu}
\nonumber 0&=&24\partial_x^4H_0+48W_1\partial_x^3H_0-3\big[24H_2-36\partial_xW_1
-9W_1^2+8m^2\big]\partial_x^2H_0\\
\nonumber &&-\,3\big[28\partial_x\big(2H_2-\partial_xW_1\big)-\big(8H_2
+30\partial_xW_1+W_1^2-8m^2\big)W_1\big]\partial_xH_0\\
\nonumber &&+\,\big[24\partial_x^3W_1-36W_1\partial_x\big(6H_2
-\partial_xW_1\big)+\big(36\partial_xW_1+11W_1^2-8m^2\big)\partial_xW_1\\ &&+\,8\big(8H_2-11\partial_xW_1+17W_1^2-8m^2\big)H_2\big]H_0+N,
\end{eqnarray}
where the non-homogeneous term, $N$, is
$$N=48\big[\partial_u\partial_x^2H_1+H_1\partial_x^2H_1+(\partial_xH_1)^2\big].$$

In summary, we have shown that the Kundt metric (\ref{Kundt}), with $W$ a polynomial function in $v$, is a solution of the vacuum equation of motion of the new theory of massive gravity (\ref{NMG}) only if $W$ is linear in $v$ and $H$ is quadratic in $v$. Moreover, each one of these Kundt solutions is locally equivalent to a metric of the form (\ref{Kundt2}), where the functions $W_1$, $H_2$, $H_1$, and $H_0$ satisfy equations (\ref{T})--(\ref{tEuu}).

Now, the scalar curvature for a Kundt metric of the form (\ref{Kundt2}) is given by
\be
\label{scurv}
R=4H_2+2\partial_xW_1-\frac{3}{2}W_1^2.
\ee
In contrast with topologically massive gravity, the scalar curvature for vacuum solutions of the new massive gravity is not constrained to take a specific value for fixed values of the parameters $\lambda$ and $m^2$. Although we will not be able to find them explicitly, there are Kundt solutions with non-constant scalar curvature. For CSI Kundt solutions of NMG, the scalar curvature will serve us to visualize how these solutions vary with $\lambda$, and $\cc$ (in one instance).

Since the equation $\tilde{T}=0$ (\ref{T}) factors nicely for $\lambda=m^2$, and since this is a special point in other respects as well, we shall isolate this special case and examine it first.

\subsection{$\lambda=m^2$}

Equation (\ref{T}) was purposefully written to make manifest the two branches of solutions when $\lambda=m^2$: we call first branch the one consisting of solutions with $H_2=W_1^2/8+m^2$, and call second branch the one consisting of solutions with $H_2=\partial_xW_1-3W_1^2/8-3m^2$. For the first branch, equation (\ref{tExx}) becomes
\be
\label{branch1}
\partial_x(W_1\Delta)-\frac{3}{2}W_1^2\Delta=0,
\ee
and for the second branch, it becomes
\be
\label{branch2}
\partial_x^2\Delta-\frac{9}{4}\partial_x(W_1\Delta)
+2\big(\partial_xW_1+\frac{7}{16}W_1^2-m^2\big)\Delta=0,
\ee
where $\Delta=\partial_xW_1-W_1^2/2-4m^2$ in both equations. These expressions have the virtue of making apparent the CSI solutions. The two branches merge at the CSI solutions with $\Delta=0$, which are of type A. (The grouping of the CSI Kundt metrics into types A, B, and C is given in the discussion following equation (\ref{CSIcondition2}).) The CSI solutions which belong to the first branch but not to the second branch are of type B, while the remaining CSI solutions (i.e., those which belong to the second branch but not to the first branch) are of type C. All these CSI solutions are included in the discussion of their counterparts at generic $\lambda/m^2$ in the following subsection. For the type C solutions, which exist only at $\lambda=m^2$, the field equations are particularly simple: $W_1=0$, $H_2=m^2$, $\partial_x^3H_1=0$, and
$$\partial_x^4H_0-4m^2\partial_x^2H_0=-2\big[\partial_u\partial_x^2H_1
+H_1\partial_x^2H_1+(\partial_xH_1)^2\big].$$
A computation of the Cotton tensor reveals that the type C solutions include non-conformally flat as well as conformally flat solutions. As remarked before, these conformally flat solutions are not Einstein. The conformally flat solutions of type C are given by: $W_1=0$, $H_2=m^2$, $\partial_x^2H_1=0$, and
$$4m^2H_0=2\partial_uH_1+H_1^2+f(u),$$
where $f$ is an arbitrary function.

In addition to the CSI solutions, both branches of solutions contain non-CSI solutions. Unfortunately, both equations (\ref{branch1}) and (\ref{branch2}) admit only translation with respect to the independent variable as point symmetry. While this means we cannot solve them directly by the group method, the translational symmetry allows us to reduce their order by one. For the first branch this reduction leads to an Abel ODE.

Finally, note that the scalar curvature for the first branch is
$$R=2\big(\partial_xW_1-\frac{1}{2}W_1^2\big)+4m^2,$$
and for the second branch, it is
$$R=6\big(\partial_xW_1-\frac{1}{2}W_1^2\big)-12m^2.$$
Since all solutions with $\partial_xW_1-W_1^2/2={\rm constant}$ are necessarily CSI, we see that the scalar curvature is constant if and only if the solution is CSI. This last statement can be strengthened and generalized. Indeed, it can be shown that the original Kundt metric (\ref{Kundt}), without the assumption of $W$ being polynomial in $v$, is a solution of NMG (\ref{NMG}) with constant scalar curvature only if $W$ is linear in $v$, and consequently it belongs (up to coordinate transformations) to the subfamily given by (\ref{Kundt2}). Furthermore, by examining  equations (\ref{scurv}) and (\ref{T}) (for general $\lambda$), it follows that $R$ is constant if and only if the Kundt solution is CSI.

\subsection{CSI solutions of NMG}

\begin{figure}[thp]
\centering
\includegraphics[width=\textwidth]{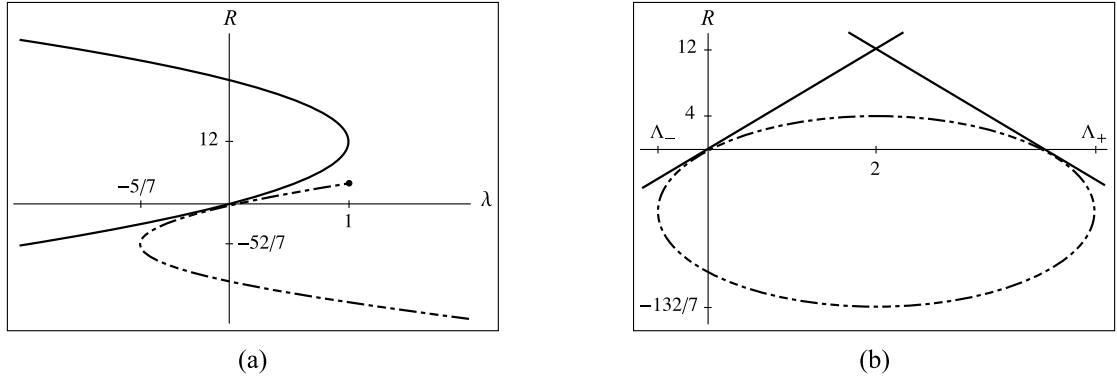}
\caption{The scalar curvature of the CSI Kundt solutions of NMG versus (a) $\lambda$ and (b) $\cc$, for positive $m^2$. The solid curves correspond to metrics of type A and the dashed curves to metrics of type B. In (a), the two parabolas meet tangentially at the point $(-35/289,-12/17)$. In (b), $\cc_\pm=2\pm4\sqrt{3/7}$, and the lines meet the ellipse tangentially at the points $(-2/17,-12/17)$ and $(70/17,-12/17)$. The three quantities $R$, $\cc$, and $\lambda$ are all expressed in units of $m^2$.}\label{NMGfig1}
\end{figure}

\begin{figure}[thp]
\centering
\includegraphics[width=\textwidth]{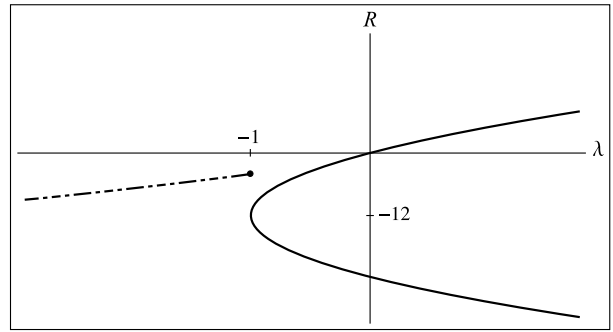}
\caption{The scalar curvature of the CSI Kundt solutions of NMG versus $\lambda$, for negative $m^2$. The solid parabola represents metrics of type A, and the dashed curve (which is part of a parabola) represents metrics of type B. $R$ and $\lambda$ are expressed in units of $|m^2|$.}\label{NMGfig2}
\end{figure}

For generic $\lambda/m^2$, in order to simplify the system of equations $\tilde{T}=0$ and $\tilde{E}_{xx}=0$ (equations (\ref{T}) and (\ref{tExx}), respectively) we \emph{assume} that the functions $H_2$ and $W_1$ are related by
$$H_2=\frac{1}{8}(W_1^2+4c_1),$$
for some constant $c_1$. Note that this restriction does not preclude any CSI solution. Conversely, it turns out that the solutions of NMG satisfying this condition are necessarily CSI. Furthermore, the system of equations (\ref{T}), (\ref{tExx}), and $H_2=(W_1^2+4c_1)/8$ admits a solution only if
$${\rm (A)}\;\;\;\;\lambda=\frac{1}{4m^2}c_1(4m^2-c_1)\;\;\;\;{\rm and}\;\;\;\;\partial_xW_1-\frac{1}{2}W_1^2=2c_1,$$
or
$${\rm (B)}\;\;\;\;\lambda=\frac{1}{108m^2}c_1(7c_1^2+44c_1m^2-8m^4)\;\;\;\;{\rm and}\;\;\;\;W_1^2=\frac{2}{9}(2m^2-c_1).$$
The NMG solutions satisfying (A) are CSI of type A, and the solutions satisfying (B) are CSI of type B, provided $c_1\neq -2m^2/17$ and $c_1\neq2m^2$. The two types of solutions merge, for positive $m^2$, at $c_1=-2m^2/17$ (with $\lambda=-35m^2/289$), where the solutions of type B disappear; this special point is the counterpart of the special point $\cc=-\m^2/9$ of TMG (see figure \ref{TMGfig1}). While the scalar curvature for the solutions of type A is $R=6c_1$, the scalar curvature of the type B solutions is $R=4(5c_1-m^2)/9$. Plots of $R$ versus $\lambda$ are given in figures \ref{NMGfig1}(a) and \ref{NMGfig2}, for positive and negative $m^2$ respectively. Solutions of type A exist only for $\lambda/m^2\leq 1$, while solutions of type B must satisfy $R\leq 4m^2$ (for $W_1$ to be real). The last condition sets an upper limit on the scalar curvature of CSI Kundt solutions of type B; this limit is realized by the CSI solutions at the special point $(\lambda=m^2,R=4m^2)$, which are of type C. For $m^2<0$, it is clear from figure \ref{NMGfig2} that CSI Kundt solutions of type A and CSI Kundt solutions of type B do not coexist. As for $m^2>0$, CSI Kundt solutions of type B exist only for $\lambda\geq -5m^2/7$ (see figure \ref{NMGfig1}(a)).

When $\lambda/m^2\leq 1$, we have seen in the preliminaries section that the vacuum equation of NMG admits a constant curvature solution, satisfying $R_{\alpha\beta}=2\cc g_{\alpha\beta}$, where $\cc$ is given by $\lambda=\cc-\cc^2/(4m^2)$. A plot of the scalar curvature of the CSI Kundt solutions versus $\cc$, for positive $m^2$, is given in figure \ref{NMGfig1}(b). The corresponding plot for negative $m^2$ consists simply of two straight lines representing solutions of type A and one point at $(\cc=2m^2,R=4m^2)$, so it is not included here.

Now, by equation (\ref{T}), when $W_1$ is constant, $H_2$ is also constant. Thus the solutions with constant $W_1$ are necessarily CSI. As is the case for TMG, the subfamily of CSI Kundt solutions of NMG with constant $W_1$ is particularly easy to describe, so we will restrict attention to these solutions in the remainder of this subsection. Equations (\ref{tEux}) and (\ref{tEuu}) are (for constant $u$) linear ODEs with constant coefficients. Thus their solutions are determined by the roots of their characteristic polynomials (and the non-homogeneous term, for equation (\ref{tEuu})).

\subsubsection{Type A solutions with constant $W_1$}

First, note that the CSI Kundt solutions of type A with constant $W_1$ exist only for $R\leq 0$, since for these solutions $W_1^2=-2R/3$. The characteristic equation of the linear ODE in $H_1$, equation (\ref{tEux}), is
$$r\big(8r^2-8W_1r+W_1^2-8m^2\big)=0,$$
with roots $0$ and $\big(W_1\pm\sqrt{4m^2+W_1^2/2}\big)/2$. For positive $m^2$, the roots are always real, and a multiple root occurs only at the special point $(\lambda=-3m^2,R=-12m^2)$, where $0$ is a double root. For negative $m^2$, the roots are all real for $R\leq12m^2$, and there are two non-real (complex conjugate) roots in the interval $12m^2<R\leq0$. In this case, all roots are distinct, except at the special point $(\lambda=m^2,R=12m^2)$, where $W_1/2$ is a double root.

Now, the characteristic equation of the linear ODE in $H_0$, equation (\ref{tEuu}), is
\be
\label{AEuu}
r(r+W_1)\big(8r^2+8W_1r+W_1^2-8m^2\big)=0,
\ee
with roots $0$, $-W_1$, and $-W_1/2\pm\sqrt{2(8m^2+W_1^2)}/4$. For positive $m^2$, the roots are always real, and multiple roots occur at the special points $(\lambda=-3m^2,R=-12m^2)$ and $(\lambda=0,R=0)$. While $0$ and $-W_1$ are both double roots at the point with $\lambda=-3m^2$, the point with $\lambda=0$ admits $0$ as its only double root. For negative $m^2$, the roots are all real for $R\leq12m^2$ (see figure \ref{NMGfig4}), and there are two non-real roots in the interval $12m^2<R\leq0$; note that these intervals match those for the characteristic polynomial of the ODE in $H_1$. All roots are distinct, except at the special points $(\lambda=m^2,R=12m^2)$ and $(\lambda=0,R=0)$. The point with $\lambda=-3m^2$ admits the double root $-W_1/2$, whereas the point with $\lambda=0$ admits $0$ as a double root.

\subsubsection{Type B solutions}

As noted earlier, the CSI Kundt solutions of type B exist only for $R\leq 4m^2$ (see figure \ref{NMGfig1}(a)), since for these solutions $W_1^2=(4m^2-R)/10$. The characteristic equation of the linear ODE in $H_1$, equation (\ref{tEux}), is
$$r(r+W_1)(r-2W_1)=0,$$
with roots $0$, $-W_1$, and $2W_1$, which are all real. A multiple root occurs only at the special point $(\lambda=m^2,R=4m^2)$, where $0$ is a triple root.

\begin{figure}[thp]
\centering
\includegraphics[width=\textwidth]{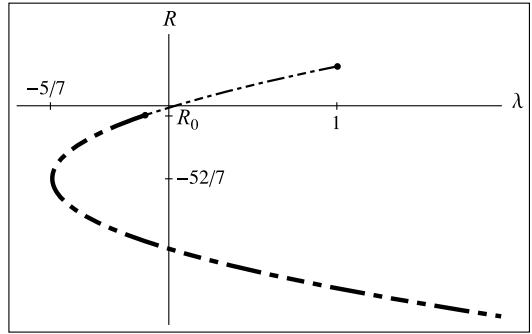}
\caption{Nature of the roots of equation (\ref{BEuu}) for the CSI Kundt solutions of NMG of type B, for $m^2>0$. The thick dashed curve indicates the presence of two non-real roots.}\label{NMGfig3}
\end{figure}

\begin{figure}[thp]
\centering
\includegraphics[width=\textwidth]{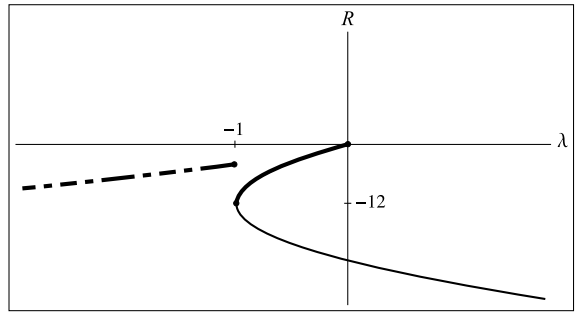}
\caption{Nature of the roots of equations (\ref{AEuu}) and (\ref{BEuu}) for $m^2<0$. The thick curves indicate the presence of two non-real roots.}\label{NMGfig4}
\end{figure}

The characteristic equation of the linear ODE in $H_0$, equation (\ref{tEuu}), is
\be
\label{BEuu}
r\big(2r^3+4W_1r^2+15W_1^2r-8m^2r-4W_1^3\big)=0.
\ee
This equation admits at least two real roots, one of which is $0$. The discriminant of the cubic factor (up to a positive multiple) is $4093R^3+5844m^2R^2+31984m^4R+28608m^6$, which admits only one real root: $R_0\approx -0.950m^2$. (Note that, for positive $m^2$, $R_0$ is smaller than the scalar curvature of the point where the CSI Kundt solutions of types A and B merge.) Thus, for positive $m^2$, the roots of equation (\ref{BEuu}) are all real in the range $R_0\leq R\leq 4m^2$, and there are two non-real roots for $R<R_0$ (see figure \ref{NMGfig3}). Multiple roots occur at $R=4m^2$ (where $0$ is a double root) and at $R=R_0$ (where the double root is $\approx -1.465 W_1$). For negative $m^2$, there is always one pair of non-real roots (see figure \ref{NMGfig4}).

\subsection{Kundt solutions of TMG which are also solutions of NMG}

Assume that the Kundt metric (\ref{Kundt2}) is a solution of the the equation of motion of TMG (\ref{TMG}). This means that the functions $W_1$, $H_2$, $H_1$, and $H_0$ satisfy equations (\ref{H2TMG})--(\ref{Euu}). In what follows, we will identify those metrics which are also solutions of the field equation of NMG (\ref{NMG}), with $\lambda=\cc-\cc^2/(4m^2)$.

With the above assumptions, equation (\ref{T}) becomes
$$\partial_xW_1-\frac{1}{2}W_1^2=2\cc,$$
which implies that the metrics must be CSI of type A. Equation (\ref{tExx}) is then automatically satisfied, whereas equations (\ref{tEux}) and (\ref{tEuu}) become
$$(2m^2-2\m^2-\cc)\partial_xH_1=0,$$
$$4(\partial_xH_1)^2-(2m^2-2\m^2-\cc)[2\partial_x^2H_0
+2W_1\partial_xH_0+(W_1^2+4\cc)H_0]=0,$$
respectively. Thus, there are two cases, depending on whether or not the parameters in the equations of motion of the two theories of massive gravity are constrained to satisfy the additional relation $2m^2=2\m^2+\cc$.

If $2m^2\neq 2\m^2+\cc$, then the Kundt metric given by (\ref{Kundt2}) is a solution to both equation (\ref{TMG}) and equation (\ref{NMG}) if and only if the following equations are satisfied
$$\partial_xW_1-\frac{1}{2}W_1^2=2\cc,\;\;\;\;H_2=\frac{1}{8}\big(W_1^2+4\cc\big),$$
$$\partial_xH_1=0,\;\;\;\;2\partial_x^2H_0+2W_1\partial_xH_0+\big(W_1^2+4\cc\big)H_0=0.$$
These are, expectedly, precisely the necessary and sufficient conditions for the Kundt metric (\ref{Kundt2}) to be a solution of Einstein's equation $R_{\alpha\beta}=2\cc g_{\alpha\beta}$.

If $2m^2=2\m^2+\cc$, then the Kundt metric given by (\ref{Kundt2}) is a solution to both equation (\ref{TMG}) and equation (\ref{NMG}) if and only if the following equations are satisfied
$$\partial_xW_1-\frac{1}{2}W_1^2=2\cc,\;\;\;\;H_2=\frac{1}{8}\big(W_1^2+4\cc\big),\;\;\;\;\partial_xH_1=0,$$
and equation (\ref{Euu}). And, for fixed $W_1$, these are precisely the necessary and sufficient conditions for the Kundt metric (\ref{Kundt2}) to be a solution of both the full and linearized equation of motion of TMG, where the linearization is with respect to the background metric
$$\bar{g}_{\alpha\beta}dx^\alpha dx^\beta=2\left[\frac{1}{8}v^2\big(W_1^2+4\cc\big)du^2+dudv
+vW_1dudx\right]+dx^2.$$

\subsection{Kundt solutions of NMG which are also solutions of linearized NMG}

We now identify, among the family of Kundt metrics (\ref{Kundt2}), a subfamily of metrics which are simultaneously solutions of the full non-linear vacuum equation of motion of the new theory of massive gravity (\ref{NMG}) and solutions of the linearized equation of motion (\ref{LNMG}), with background metric (\ref{background}).

Let's assume the Kundt metric given by (\ref{Kundt2}) is a solution of the full equation of motion (\ref{NMG}) and that $W_1=\bar{W}_1$. Then equation (\ref{T}) becomes
$$(8H_2-\bar{W}_1^2-4\cc)(8H_2-\bar{W}_1^2+16m^2-12\cc)=0,$$
and $R^{(1)}$ is given by
$$R^{(1)}=\frac{1}{2}(8H_2-\bar{W}_1^2-4\cc).$$
Recall that for $\cc\neq 2m^2$, $R^{(1)}$ must be identically zero. Thus, whether $\cc$ is equal to $2m^2$ or not, the last two equations are satisfied if and only if
$$H_2=\frac{1}{8}\big(\bar{W}_1^2+4\cc\big).$$

Now let $\tilde{L}_{\alpha\beta}$ denote the left-hand side of the linearized equation of motion (\ref{LNMG}):
$$\tilde{L}_{\alpha\beta}=R^{(1)}_{\alpha\beta}
-\frac{1}{2}R^{(1)}\bar{g}_{\alpha\beta}+(\lambda-3\cc) h_{\alpha\beta}-\frac{1}{2m^2}K^{(1)}_{\alpha\beta}.$$
A direct computation of $\tilde{L}_{\alpha\beta}$ shows that all components are identically zero (modulo the full nonlinear field equations), except $\tilde{L}_{uu}$, which gives the additional equation
$$\partial_x\big(H_1\partial_xH_1\big)=0.$$
This last equation and equation (\ref{tEux}) are satisfied simultaneously if and only if $\partial_xH_1=0$.
In summary, we have shown that the Kundt metrics (\ref{Kundt2}), satisfying the following equations
$$W_1=\bar{W}_1,\;\;\;\; H_2=\frac{1}{8}\big(\bar{W}_1^2+4\cc\big),\;\;\;\;\partial_xH_1=0,$$
and equation (\ref{tEuu}), are solutions of both the full and linearized equation of motion of NMG, where the linearization is with respect to the background metric (\ref{background}).

\section{Conclusion}

We have studied the problem of which Kundt metrics are local solutions of two theories of massive gravity in three dimensions: topologically massive gravity and the new theory of massive gravity, proposed recently in \cite{Bergshoeff:2009hq}. For topologically massive gravity, this question was considered, very recently, in \cite{Chow}. Due to the existence of a local coordinate system in which a Kundt metric takes the form (\ref{Kundt}), the field equations of TMG reduce for this class of metrics to a simple set of three ODEs: one non-linear second-order ODE in $W_1$, equation (\ref{Exx}), and two linear ODEs, equations (\ref{Eux}) and (\ref{Euu}), which are of order two and three, respectively. Only the CSI solutions (i.e., solutions with constant scalar polynomial curvature invariants) were found explicitly in \cite{Chow}. An application of Lie's theory of differential equations to the non-linear ODE leads to the first non-CSI explicit solutions of TMG, and remarkably this happens only when $\cc=-\m^2$. With a choice of appropriate constant-curvature backgrounds, we characterized, among the family of Kundt metrics (\ref{Kundt2}), those which are solutions of both the full and linearized equations of motion of TMG.

For the new theory of massive gravity, the field equations for the Kundt metric (\ref{Kundt}), with the extra assumption that $W$ is polynomial in the null coordinate $v$, reduce to a simple set of four ODEs: equations (\ref{T})--(\ref{tEuu}). The CSI solutions were discussed in detail, with focus on the subfamily for which equations (\ref{tEux}) and (\ref{tEuu}) are linear ODEs with constant coefficients. These solutions were classified according to the Segre types of their Ricci and Cotton tensors and by their scalar curvature. With the same choice of backgrounds as for TMG, a family of Kundt solutions of NMG which are also solutions of the linearized theory was found. Furthermore, we have also answered a question about which Kundt solutions of TMG are also solutions of NMG.

A number of plots of either the square of the Ricci tensor (for TMG) or the scalar curvature (for NMG) versus either $\cc$ or $\lambda$, in units of the explicit mass parameters ($\m$ for TMG and $m$ for NMG), were provided. Besides providing a panoramic view of the space of CSI solutions across all values of $\cc$ and $\lambda$ (in units of the explicit mass parameters), they make visible the special nature of some values of $\cc$ and $\lambda$. For instance, the point $(\lambda=m^2,R=4m^2)$ in figures \ref{NMGfig1}(a) and \ref{NMGfig2} is the only place where conformally flat, non-Einstein, CSI Kundt solutions of NMG exist. Incidentally, this interesting point also includes the static Einstein universe $\mathbb{R}\times \mathrm{S}^2$ (for positive $m^2$) and the static universe with negative spatial curvature $\mathbb{R}\times \mathrm{H}^2$ (for negative $m^2$) \cite{Oliva}, which are non-Kundt vacuum solutions of NMG; the sphere $\mathrm{S}^2$ and the hyperbolic plane $\mathrm{H}^2$ both have radius $1/(2|m^2|)$. Also, the point $\cc=-\m^2/9$ of figure \ref{TMGfig1} and the point $(\lambda=-35m^2/289,R=-12m^2/17)$ of figure \ref{NMGfig1}(a) are analogous in that the neighboring CSI Kundt solutions of type B disappear at these points.

Finally, we note that subfamilies of the CSI Kundt vacuum solutions of massive gravity in three dimensions presented here and in \cite{Chow} have already been found previously in the literature. Solutions of TMG of this type are discussed in detail in \cite{Chow}. For NMG, the AdS wave solutions of \cite{Ayon} belong to the family of metrics given by (\ref{Kundt2}), with $H_2=H_1=W_1=0$, and the black holes of \cite{Clement} are special cases of the AdS waves of \cite{Ayon}.

\section*{Acknowledgements}

I would like to thank Jacques Distler for reading the manuscript and making helpful suggestions. This work was supported by NSF grant PHY-0455649.

\end{document}